\documentclass[a4paper,12pt]{article}
\usepackage[left=0.8in,right=0.8in,top=0.8in,bottom=0.8in]{geometry}
\usepackage{amssymb}
\usepackage{amsmath}
\usepackage{physics}
\usepackage{graphicx}
\usepackage{float}
\usepackage{slashed}
\usepackage{appendix}
\usepackage{bm}
\usepackage[hidelinks]{hyperref}
\usepackage[sorting=none, style = phys, biblabel=brackets, backend = biber]{biblatex}
\usepackage{authblk}
\newcommand\blfootnote[1]{%
  \begingroup
  \renewcommand\thefootnote{}\footnote{#1}%
  \addtocounter{footnote}{-1}%
  \endgroup
}
\DeclareUnicodeCharacter{0301}{\'{i}}
\bibliography{A_SC_for_spin_chains}

\newcommand{\be}{\begin{equation}}
\newcommand{\ee}{\end{equation}}
\newcommand{\bea}{\begin{eqnarray}}
\newcommand{\eea}{\end{eqnarray}}

\title{\textbf{Spread complexity and quantum chaos for periodically driven spin chains}}
\author{Amin A. Nizami$^{*}$}
\author{Ankit W. Shrestha$^{\dagger}$}
 \affil{\small{Department of Physics, Ashoka University, Rajiv Gandhi Education City, Rai, NCR, India 131029}}
\date{}

\begin{document}

\maketitle
\begin{abstract}
The complexity of quantum states under dynamical evolution can be investigated by studying the spread with time of the state over a pre-defined basis. It is known that this complexity is minimised by choosing the Krylov basis, thus defining the spread complexity. We study the dynamics of spread complexity for quantum maps using the Arnoldi iterative procedure. The main illustrative quantum many-body model we use  is the periodically kicked Ising spin-chain with non-integrable deformations, a chaotic system where we look at both local and non-local interactions. In the various cases we find distinctive behaviour of the Arnoldi coefficients and spread complexity for regular vs. chaotic dynamics: suppressed fluctuations in the Arnoldi coefficients as well as larger saturation value in spread complexity in the chaotic case. We compare the behaviour of the Krylov measures with that of standard spectral diagnostics of chaos. We also study the effect of changing the driving frequency on the complexity saturation. 
    
    \blfootnote{ $^*$amin.nizami@ashoka.edu.in  \hspace{0.6cm} $^\dagger$sth.ankit61@gmail.com }
\end{abstract}

\section{Introduction}
The complexity of physical systems in general, and for quantum dynamics of many-body systems in particular, is encapsulated quantitatively by a number of measures. These include circuit complexity, quantum Kolmogorov complexity, algorithmic randomness, entanglement based measures and entropic measures \cite{Nielsen:2005mkt, Nielsen:2006cea, Dowling:2006tnk, Zurek:1989zz, BERTHIAUME2001201}. The detailed dynamics of complexity can also signal the presence (or absence) of quantum chaos, and thus these complexity measures can often be used as chaos diagnostics \cite{Ali:2019zcj, Cotler:2017jue, Roberts:2016hpo}. 

A relatively new measure of the complexity of operator growth with bearing on the integrable or chaotic nature of the dynamics is Krylov complexity (K-complexity).  
Operator K-complexity was first defined and studied in \cite{Parker_2019} for quantum systems in the thermodynamic limit. Its relation with circuit complexity was recently studied in \cite{Craps_2024, Lv:2023jbv, Aguilar-Gutierrez:2023nyk}. It utilises the Heisenberg picture of quantum dynamics and measures the growth of operator complexity as encapsulated by successive nested commutators of the system Hamiltonian with an (initially localised) observable.  For maximally chaotic systems, the initial growth rate (upto the scrambling time) of K-complexity is exponential \footnote{However, this can happen also for cases involving saddle-dominated scrambling resulting (semi-classically) from unstable fixed points in phase space \cite{Bhattacharjee:2022vlt}. Also for CFTs, even rational and free ones, the complexity growth is exponential \cite{Dymarsky:2021bjq}, see also \cite{Chattopadhyay:2024pdj}. So this is not a sufficient condition for chaos.}. For finite size systems this growth eventually becomes linear and saturates after the Heisenberg time \cite{Rabinovici:2019wsy}. The universal operator growth hypothesis \cite{Parker_2019} posits a linear growth of the Lanczos coefficients (defined in the next section) for the maximally chaotic case whereas their growth is sub-linear for the integrable cases. For free (quadratic) models, they are typically constant (their dependence on operator size is explored in \cite{Kim:2021okd}). For a comprehensive recent review on Krylov construction in quantum systems, see \cite{Nandy:2024htc}.

The main focus of this paper is the related notion of K-complexity for quantum states, also known as state complexity or spread complexity. It was first introduced in \cite{Balasubramanian_2022} and utilises the Schr\"{o}dinger picture of state evolution. It measures the spread with time of an initial state over a pre-defined basis, namely the Krylov basis that is defined iteratively through the Lanczos algorithm. The next section reviews the basic definitions and concepts relating to spread complexity that we will require. Spread complexity has been investigated for a variety of models starting from the Sachdev-Ye-Kitaev (SYK) model and Random Matrix Models in \cite{Balasubramanian_2022}. It has been studied for spin chains \cite{Gill:2023umm, Scialchi:2023bmw}, the quenched LMG model \cite{Afrasiar:2022efk, Bento:2023bjn}, Kitaev chain \cite{Caputa:2022yju}, free fermion models \cite{Gautam:2023pny}, many-body scars \cite{Bhattacharjee:2022qjw, Nandy:2023brt}, and others \cite{Chattopadhyay:2023fob, Pal:2023yik, Gautam:2023bcm, Bhattacharya:2023yec, Dixit:2023fke, Huh:2023jxt, Carolan:2024wov, Balasubramanian:2023kwd, Zhou:2024rtg, Erdmenger:2023wjg, Bhattacharya:2023zqt}. It also serves as an order parameter which can discern topologically non-trivial phases of matter \cite{Caputa:2022eye}.

In this paper we will study spread complexity for quantum maps such as those generated by Floquet dynamics of spin-chains. Precision Floquet engineering in periodically driven quantum systems opens up many avenues, both experimental and theoretical, in  quantum control using optical lattices and ultra-cold atoms \cite{Eckardt:2016lof, Weitenberg_2021, Meinert:2016}. It can be used to study novel non-equilibrium phenomenon and engineer novel phases of matter such as time crystals \cite{Zaletel:2023aej}. 


We will study chaotic dynamics in a quantum many-body system -- the periodically kicked Ising spin-chain. Such a lattice of interacting spin-half fermions provides an ideal testing ground for many-body quantum chaos. As this intrinsically quantum mechanical system has no classical limit, it is useful in exploring facets of chaotic dynamics beyond the semi-classical regime. A primary objective in our study is to investigate the transition from regular to maximally chaotic dynamics in Floquet spin-chain models when the coupling strength or the non-locality of the interaction is tuned. This is done using a variety of measures, with a particular focus on spread complexity. The first model here will be the kicked Ising spin-chain with a tunable coupling constant that interpolates between the integrable and chaotic regimes. We will also look at a maximally non-local version of the Kicked Ising spin chain and study the effect of non-locality on the dynamics.
Maximally non-local spin-chain models with constant all-to-all interactions have been studied earlier as systems exhibiting fast scrambling \cite{Belyansky:2020bia, Pappalardi:2018frz}. Fast scramblers have a scrambling time $t_s \sim \log N$ where $N$ is the system size and have maximally chaotic dynamics with the quantum Lyapunov exponent saturating the chaos bound \cite{Maldacena:2015waa}. The second model (discussed in appendix C) is a bosonic spin-chain: the kicked Bose-Hubbard dimer model.


This paper is organised as follows. Section \ref{rev} introduces and reviews  the basic notions of spread complexity and the Arnoldi iterative algorithm utilised for Krylov construction for quantum maps. We also review the Krylov chain picture and use it to give a simple argument for the complexity growth being at most linear in time. 
Section \ref{ising} introduces the kicked Ising spin chain model, both the local and non-local versions. Here  we also study the distinctive behaviour in the integrable and chaotic domain, of the Arnoldi coefficients, spread complexity, as well as other chaos quantifiers such as the spectral parameter $\eta$ and the delocalisation parameter $\sigma$. Some similar results for the Bose-Hubbard dimer model are relegated to Appendix C. In section \ref{dis} 
we conclude with a discussion of the main results and observations. 

\section{Arnoldi iteration and spread complexity for quantum maps} \label{rev}
In this section we review the basic notions of Krylov construction, spread complexity, Lanczos and Arnoldi iteration that we will need in our subsequent analysis. For more details, the reader is referred to the introductory sections of \cite{Balasubramanian_2022, Parker_2019, geometry, Rabinovici_2021, Nizami:2023dkf} and the review \cite{Nandy:2024htc}.  

\subsection{Krylov complexity for quantum states}
Consider a basis $\mathfrak{B}=\{ \ket{B_n}\}$ using which a time-dependent quantum state is expanded as $\ket{\psi(t)}=\sum_n \psi_n(t)\ket{B_n}$. Following \cite{Balasubramanian_2022} we define a complexity measure
\begin{align}
C_{\mathfrak{B}}(t)=\sum_n n \, |\psi_n(t)|^ 2 \label{cost_function}
\end{align}
which quantifies the (time-dependent) spread of the state over the basis elements. A functional minimization over the bases: $C(t)=min_{\mathfrak{B}}\,\, C_{\mathfrak{B}}(t)$, leads to a basis-independent measure of state complexity dubbed as spread complexity (see \cite{Balasubramanian_2022} for more details). These authors also showed that this minimum is obtained for the Krylov basis. The Krylov construction for quantum states for autonomous as well as Floquet systems is outlined below.

\subsection{Review of Lanczos iteration}
\begin{align}
\ket{\psi(t)}=\exp (-itH) \ket{\psi_0}=\sum_{n=0}^ {\infty} \frac{(-it)^ n}{n!}H^n \ket{\psi_0},
\end{align}
which suggests the natural basis $\{\ket{\psi_0}, H\ket{\psi_0}, H^2\ket{\psi_0}, \dots\}$ in which the state can be expanded. We assume that the initial state is not an eigenstate of the full Hamiltonian. Krylov construction then proceeds by orthonormalising this set using the Lanczos algorithm.
The algorithm iteratively generates a set of orthonormal basis vectors starting from $\ket{K_{-1}} \equiv 0$, $\ket{K_0} = \ket{\psi_0}$ and 
\begin{align}
    \ket{K_n} = \frac{1}{b_n} \ket{A_n},  \;\;\;\;\;\;\ket{A_{n+1}} = (H - a_n)\ket{K_n} -b_n\ket{K_{n-1}} 
\end{align}
for $n\geq 1$. The Lanczos coefficients are defined as:
\begin{align}
    a_n = \bra{K_n}H \ket{K_n},\;\;\; b_n = \braket{A_n}^{1/2}
\end{align}

This algorithm naturally halts for systems with a finite-dimensional Hilbert space after generating a set of orthonormal states. The Hamiltonian in the Krylov Basis takes a tri-diagonal form with $a_n$ as the diagonal elements and $b_n$ as the sub-diagonal and super-diagonal elements \cite{Balasubramanian_2022}.  The time-evolved state $\ket{\psi(t)}$ can be expanded in the Krylov Basis as:
\begin{align}
    \ket{\psi(t)} = \sum_{n = 0}^{D_K-1} \psi_n(t) \ket{K_n}
\end{align}
where $D_K$ is the dimension of the Krylov space. The total probability $\sum_n |\psi_n(t)|^2 = 1$  is conserved for unitary evolution.  Spread complexity is then defined as:
\begin{align}
    C(t) = \sum_{n=0}^{D_K-1} n |\psi_n(t)|^2
\end{align}
which measures the average spread of the wave-function over the Krylov Basis \cite{Balasubramanian_2022}. An equivalent form of  spread complexity is the expectation value of the Krylov complexity operator $\hat{K}$:
\begin{align}
    C(t) = \bra{\psi(t)}\hat{K}\ket{\psi(t)}, \;\;\;\;\; \hat{K} = \sum_{n=0}^{D_K-1}n\ket{K_n}\bra{K_n}
\end{align} 

Apart from the average spread, we can also study the Shannon entropy of the spread using the probability distribution $|\psi_n(t)|^2$. It is defined as
\begin{align}
    S(t) = -\sum_{n=0}^{D_K-1}|\psi_n(t)|^2\, \ln |\psi_n(t)|^2
\end{align}

\subsection{Review of Arnoldi iteration}\label{AC}
For periodically driven systems, the Krylov construction proceeds similarly using the unitary Floquet operator $U_F$ and the Arnoldi iterative algorithm which generalises the Lanczos procedure beyond the Hermitian case. Here one considers the basis which is naturally constructed by stroboscopically probing the system at equal time periods $\{\ket{\psi_0}, U_F\ket{\psi_0}, U_F^2\ket{\psi_0},...\}$, where the $n^{th}$ element of the basis represents the state after $n$ time steps. In \cite{Balasubramanian_2022} it was also shown that the basis generated in this manner minimizes the complexity function defined in eq.\eqref{cost_function}.

Arnoldi orthonormalisation entails the following iterative construction. Starting with $\ket{K_0}=\ket{\psi_0}$, we define the next basis element
\begin{align}
    &\ket{K_1} = \frac{1}{h_{1,0}} \biggl[U_F \ket{K_0} - h_{0,0} \ket{K_0} \biggr].
\end{align}
Subsequent basis vectors are generated as
\begin{align}
    &\ket{K_n} = \frac{1}{h_{n,n-1}} \left[U_F \ket{K_{n-1}} - \sum_{j=0}^{n-1} h_{j,n-1} \ket{K_j} \right] \label{UF on K},
\end{align}
for $n \geq 2$, where $h_{j,k} = \bra{K_j} U_F \ket{K_j}$ are called Arnoldi coefficients and are analogous to the Lanczos coefficients in the time-independent formulation. Of these, the normalization constants $h_{n,n-1}$ are analogous to the Lanczos $b_n$'s. 
This generalization allows the algorithm to generate orthonormal vectors using non-Hermitian operators and was first used for Floquet systems in \cite{Nizami:2023dkf, Yates:2021asz} and for open quantum systems in \cite{bhattacharya_operator_2022, Bhattacharjee:2022lzy}. The unitary Floquet operator takes the Hessenberg form in the Krylov Basis with non-zero elements $h_{j,k}$ on and above the sub-diagonal of the matrix. The Arnoldi coefficients thus define the matrix representation of the system Floquet operator in the Krylov basis.

As before, the state after $j$ time-steps can be expanded in the Krylov basis
\begin{align}
    \ket{\psi_j} = \sum_{n=0}^{D_K-1} \psi_n^{\,j} \ket{K_n} \label{basis expansion}
\end{align}
and spread complexity and entropy can be defined for Floquet dynamics
\begin{align}
    C_j = \sum_{n=0}^{D_K-1} n |\psi_n^{\,j}|^2, \;\;\;\;\;\;\;\;\; S_j = -\sum_{n=0}^{D_K-1} |\psi_n^{\,j}|^2 \, \ln|\psi_n^{\,j}|^2
\end{align}

Besides kicked systems, quantum maps representing such discrete unitary evolutions also arise naturally in studying the dynamics of quantum circuits \cite{suchsland_krylov_2023}.

\subsection{Krylov Chain picture}\label{KCP}
In the Krylov construction, the dynamics of a general quantum system is mapped to a particle-hopping tight-binding model on a one-dimensional lattice with the Lanczos coefficients representing the hopping amplitudes. The equations describing this dynamics were first derived in \cite{Parker_2019} for time-independent Hamiltonian systems. For the case of Floquet dynamics, the corresponding equations were first derived in \cite{Nizami:2023dkf} and will be relevant to us here.
We take the action of the Floquet operator on eq. \eqref{basis expansion} to get:
\begin{align}
    \ket{\psi_{j+1}} = \sum_{n=0}^{D_K-1} \psi_n^{\,j}\; U_F\ket{K_n}
\end{align}
Using the Krylov basis expansion on the left hand side and action of the Floquet operator on $\ket{K_n}$ from eq. \eqref{UF on K}, we get
a set of linear difference equations for the probability amplitudes at different time steps
\begin{align}
    \psi_n^{\,j+1} = \sum_{l=n-1}^{D_K-1}h_{n,l}\, \psi_l^{\,j} \label{difference equation}
\end{align}

\begin{figure}[H]
    \centering
    \includegraphics[height = 3.8cm, width = 10.5 cm]{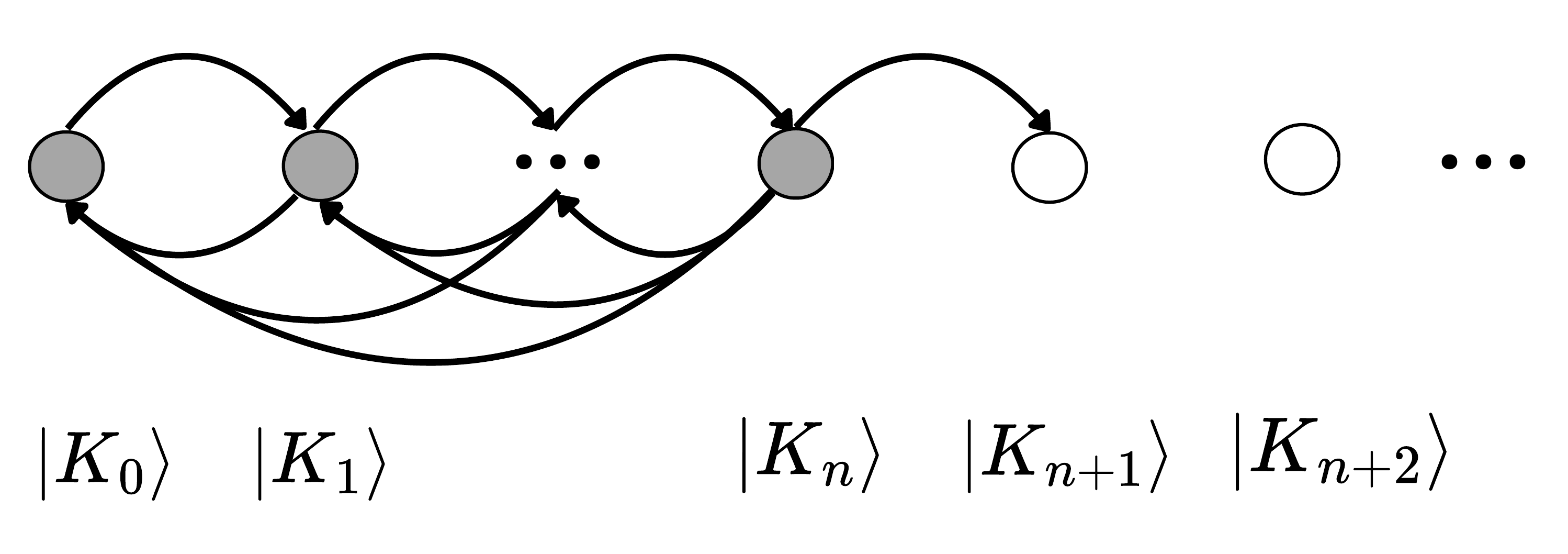}
    \caption{\small Schematic diagram of a particle hopping in a one-dimensional lattice in the Arnoldi approach to Floquet dynamics in Krylov space.  The equations \eqref{difference equation} encapsulating the dynamics imply that the particle  jumps are local to the right but can be non-local to the left, as shown in the above figure. }
    \label{fig:Krylov Chain}
\end{figure}

This equation describes a particle hopping in a one-dimensional lattice with Krylov Basis vectors as lattice sites and the coefficients $h_{n,l}$ as hopping amplitudes. In this case, the hopping is non-local in one direction (see fig.\ref{fig:Krylov Chain}) unlike the case for time-independent Hamiltonians. Solving the difference equations with the initial condition $\psi_n^0 = \delta_{n0}$ gives an equivalent way of calculating the probability amplitudes to determine the spread complexity.

This picture can be used to argue that the spread complexity growth in a Floquet system is at most linear in time, an observation made in \cite{suchsland_krylov_2023} for operator complexity growth in dual-unitary models. We will provide a more general argument. Intuitively, the idea is that since the particle on the Krylov chain can hop only one step to the right (fig.\ref{fig:Krylov Chain}), after $j$ time steps it can at most be $j$ lattice sites away to the right from its starting point. This leads to a K-complexity that is at most $j$, and hence the growth can at most be linear in time.

A more quantitative argument is as follows. The growth of spread complexity is maximised when the set of vectors generated by the repeated action of the Floquet operator for a given initial state is naturally orthonormal. In this case, eq. \eqref{UF on K} implies that the Arnoldi coefficients $h_{n,n-1}$ are equal to one and the rest of the $h_{j,k}$ are equal to zero. Eq. \eqref{difference equation} then takes the simple form
\begin{align}
    \psi_n^{\,j+1} = h_{n,n-1}\, \psi_{n-1}^{\,j} = \psi_{n-1}^{\,j}
\end{align}
Using the given initial condition, we can deduce that $\psi_n^{\,j} = \delta_{nj}$ for the case of maximum growth and spread complexity is given by
\begin{align}
    C_j^{\,max} = \sum_{n=0}^{D_K-1} n |\delta_{nj}|^2 = j
\end{align}
which gives a linear growth (with slope one) of the complexity. We will see below in the example of the spin-chains which we study that this bound on the initial growth rate is obeyed in every case. The spread entropy in such cases is zero since the particle hops from one site to the other without any dispersion in wavefunction.  

\section{Kicked Ising spin chain: local and non-local interactions} \label{ising}
In this section we will investigate facets of regular and chaotic dynamics of a quantum many-body system -- the kicked one-dimensional Ising spin-chain. This consists of a lattice of spin-half fermions with interactions that can be local (nearest-neighbour) or non-local. Besides spread complexity, we will also compute a number of other quantities and compare and contrast their behaviour. Some recent studies of complexity and chaos in (time-independent) spin chains include \cite{integrability_to_chaos, Scialchi:2023bmw, Craps:2019rbj, Craps:2023rur, Camargo:2024deu, Karthik_2007}. Aspects of quantum chaos for kicked spin-chains  have been studied recently in \cite{Herrmann:2023hdj} utilising spectral parameters and entanglement entropy.

We will begin with the local model with nearest neighbour interactions. This model consists of $N$ locally interacting spin-half systems kicked periodically with an external tilted magnetic field. The Hamiltonian for the system is \cite{Akila:2016ltx}:
\begin{align} \label{LSC}
    H &= H_0+V\, \delta_T(t) \nonumber \\ &= \sum_{i=1}^N \left(J \sigma_i^z \sigma_{i+1}^z + \bm{b.\sigma_i} \delta_T(t)\right); \,\,\, 
    \delta_T(t)=\sum_{n = -\infty }^\infty \delta(t - n T)
\end{align}
where  $\sigma_i^j$ is the Pauli matrix for the particle at site $i$ along the $j$ axis, and $\bm{b}$ is the magnetic field vector. The $z-$axis for the system can be set such that the magnetic field is along the $(x-z)$ plane and the spin-chain along the $y-$axis. The tilted magnetic field vector can be resolved into its components using an angular variable $\phi$ such that $\bm{b} = (b \sin{\phi}, 0, b\cos{\phi})$. We will mostly work with the time period $T$ set equal to one. The Floquet operator for the given Hamiltonian is given by the quantum map:
\begin{align}\label{FSC}
    U_F = \exp{-\text{i}\, J\sum_{i=1}^N\sigma_i^z \sigma_{i+1}^z} \exp{  -\text{i\,}b\sum_{i=1}^N (\sigma_i^x\,\sin{\phi}\,  + \sigma_i^z\,\cos{\phi}\, )}
\end{align}

The variable $\phi$ can be used as a tuning parameter interpolating from the integrable limit to maximal chaos. Given the symmetry of the system, we only need to study it for the range $\phi \in [0, \pi/2]$.  Besides $\phi=0$, the strongly coupled $\phi=\pi/2$ case is also integrable as it can be solved exactly by the Jordan-Wigner transformation \cite{Lieb:1961fr}. We will work with open boundary conditions, and work in a symmetry restricted system subspace, namely the positive parity sector.

\subsection{Spectral Parameter and Level Statistics}

Random Matrix Theory (RMT) and spectral statistics of many-body hamiltonians provide a classic diagnostic of chaos in the quantum domain. As per the Berry-Tabor conjecture \cite{berry1977level}, the spectral statistics of classically integrable hamiltonians is Poissonian in character. On the other hand, the BGS conjecture \cite{Bohigas:1983er} surmises that for the chaotic case, one of the several Wigner-Dyson universality classes showing level repulsion is realised. For a pedagogical reference emphasising the spectral and RMT aspects of quantum chaos, see \cite{haake_quantum_2010}. For our purposes, the integrability to chaos transition in the kicked Ising model can be characterized by the spectral statistics of the quasi-energies of the Floquet operator. Given the eigenvalue relation for the Floquet operator:
\begin{align}
    U_F \ket{\psi_\varphi} = e^{i\varphi} \ket{\psi_\varphi},
\end{align}
we can calculate the spacing between consecutive eigenvalues $s_n = \frac{N}{2\pi} (\varphi_n - \varphi_{n-1})$ and define a parameter $\langle \tilde{r} \rangle$:
\begin{align}
    \langle \tilde{r} \rangle = \frac{1}{D}\sum_{n=1}^D \frac{\min(s_n, s_{n-1})}{\max(s_n, s_{n-1})}.
\end{align}

The value of the parameter $\langle \tilde{r} \rangle$ depends on the distribution of level spacing and encodes information about the integrable or chaotic regime of the dynamics. From RMT, the analytic values of $\langle \tilde{r} \rangle$ for the Poisson and GOE cases are $\langle \tilde{r} \rangle_{P} \approx 0.38629$ and $\langle \tilde{r} \rangle_{GOE} \approx 0.53590$ \cite{Atas:2013gvn}. Thus, we can define a normalized parameter $\eta$ as:
\begin{align}\label{etadefn}
    \eta = \frac{\langle \tilde{r} \rangle - \langle \tilde{r} \rangle_{P}}{\langle \tilde{r} \rangle_{GOE}-\langle \tilde{r} \rangle_{P}}
\end{align}
The value of $\eta$ is close to zero for integrable dynamics and one for the maximally chaotic case. For the Floquet spin-chain represented by eq. \eqref{LSC} the variation of $\eta$ for different values of the coupling $\phi$ is given below. Note the smaller values of $\eta$ close to the integrable limits $\phi=0, \pi/2$. Similar computations for $\eta$ for time-independent spin-chains have been done in \cite{Scialchi:2023bmw}. Also plotted below are the spectral statistics curves for the integrable and chaotic case. This makes clear the regular nature of the dynamics for the weakly-coupled case, as the distribution of level spacings is Poissonian. For a larger value ($\phi=\pi /3$) of the coupling, the distribution shows level repulsion characterising chaos, and is of the GOE universality class.
\begin{figure}[H]
    \centering
    \includegraphics[height = 6.0 cm, width = 8.5 cm]{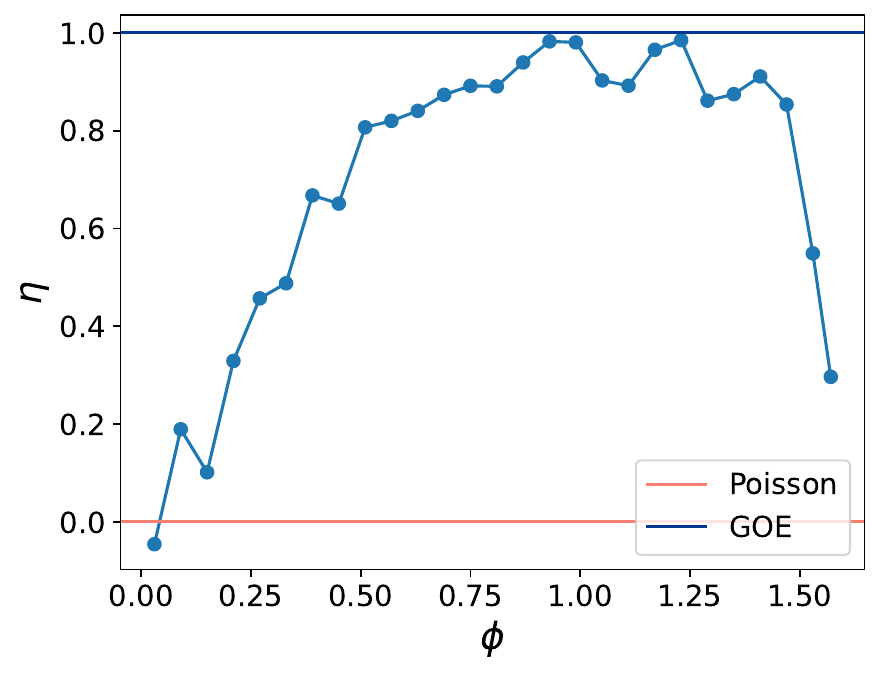}
    \caption{\small  The value of the spectral parameter $\eta$ as a function of angle $\phi$ for $N=11$ spins as the dynamics changes from regular to chaotic. Note the low value at the two ends when the system is integrable}
    \label{fig:r_stat}
\end{figure}

\begin{figure}[H]
    \centering
    \includegraphics[height = 5.7cm, width = 8.0 cm]{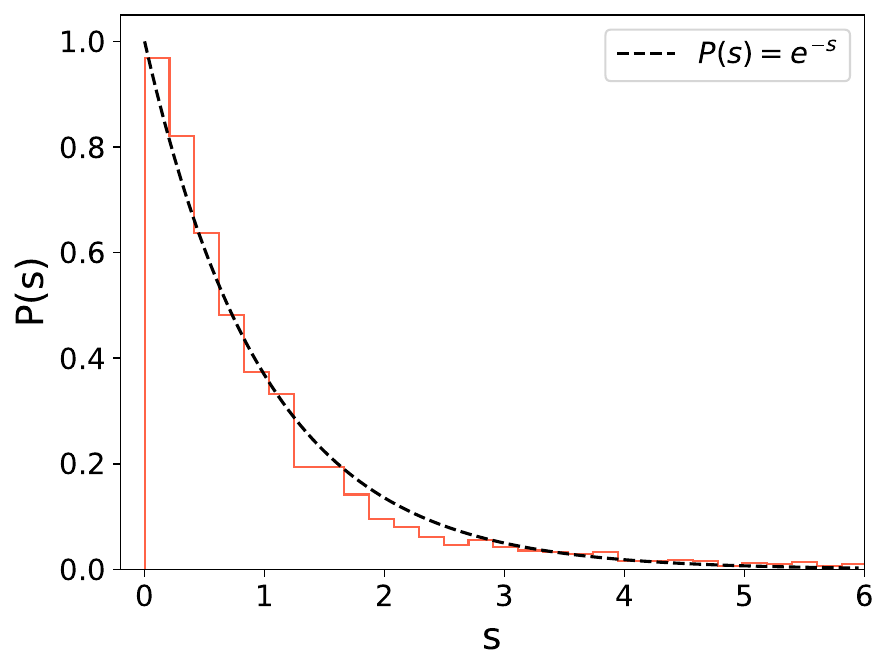}
    \includegraphics[height = 5.7 cm, width = 8.0 cm]{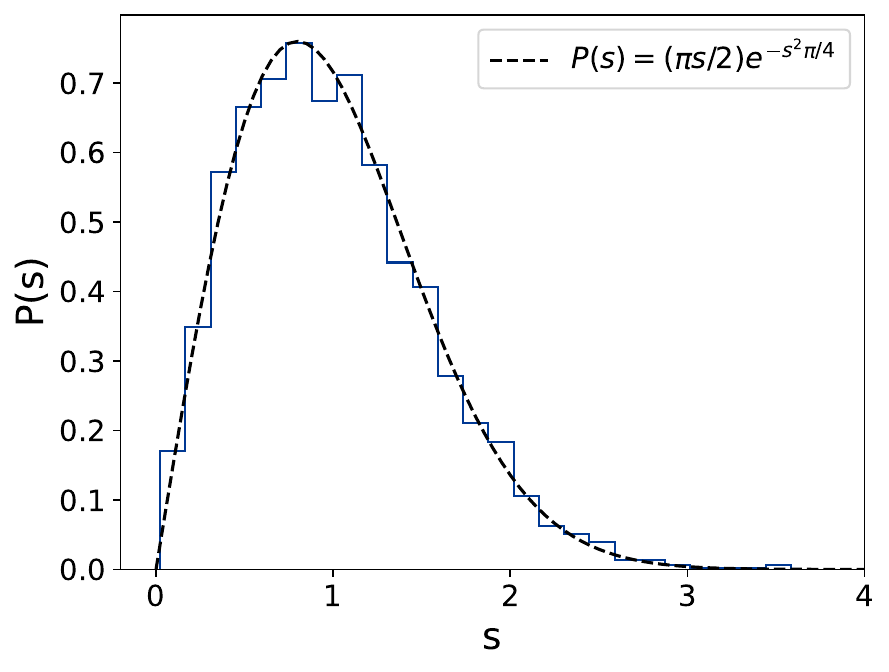}
    \caption{\small The probability distribution of the level spacing for $N=13$ spins -- Poissonian for $\phi = \pi/30$ (left) and GOE for $\phi = \pi/3$ (right). }
    \label{fig:r_stat_int_chaos}
\end{figure}

\subsection{Dynamics of Arnoldi coefficients and spread complexity}

We will now turn to features of the Arnoldi coefficients and the dynamics of spread complexity for the kicked Ising spin-chain. The various relevant quantities are defined in subsection \ref{AC}. For the following calculations, the initial state is taken as an $H_0$ eigenstate $\ket{\psi_0} = \ket{\psi_0^{eig}}$ in the positive parity sector of the Hilbert space. Most results, however, are generic with same qualitative features for other initial states. To facilitate comparison, we will keep fixed values of the coupling constants $b$ and $\phi$ for the regular as well as the chaotic case.

\begin{figure}[H]
    \centering
    \includegraphics[height = 5.8 cm, width = 8.0 cm]{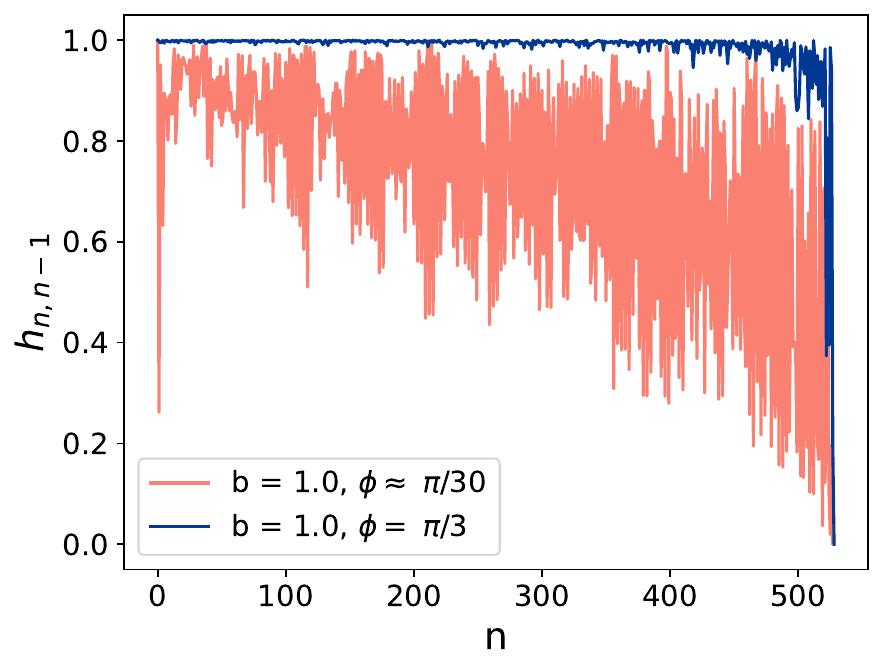}
    \includegraphics[height = 5.8 cm, width = 8.0 cm]{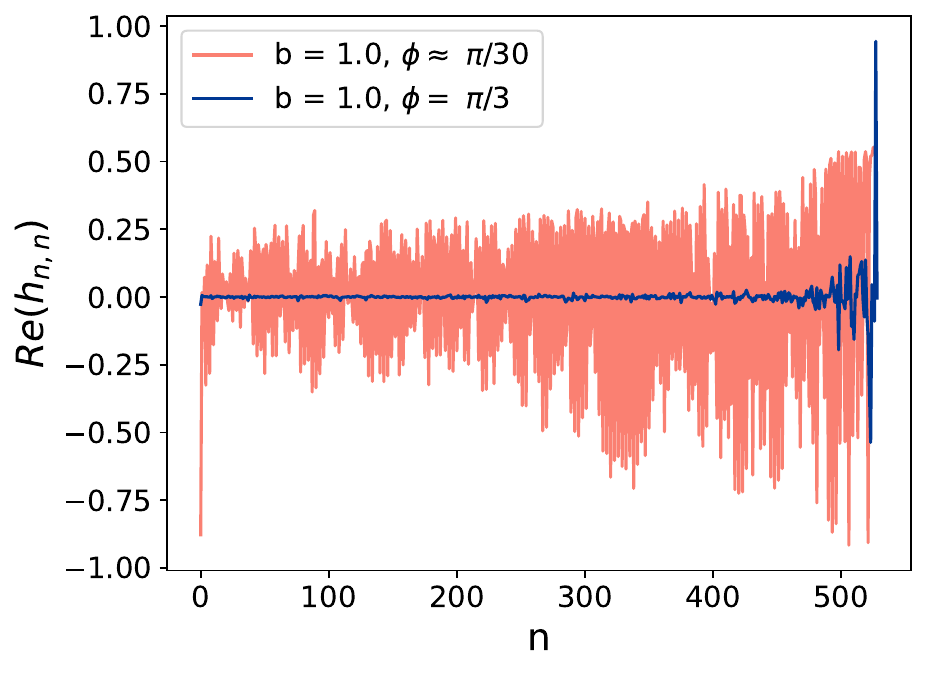}
    \caption{\small Dispersion in Arnoldi coefficients in the near integrable and chaotic limits. The left plot is for the real sub-diagonal $h_{n,n-1}$ whereas the right plot is for the real part of the diagonal $h_{n,n}$ -- the imaginary part fluctuates in a similar manner. The calculations are done for $N=10$ in the positive parity sector.}
    \label{fig:State}
\end{figure}
\begin{figure}[H]
    \centering
    \includegraphics[height = 5.8 cm, width = 8.0 cm]{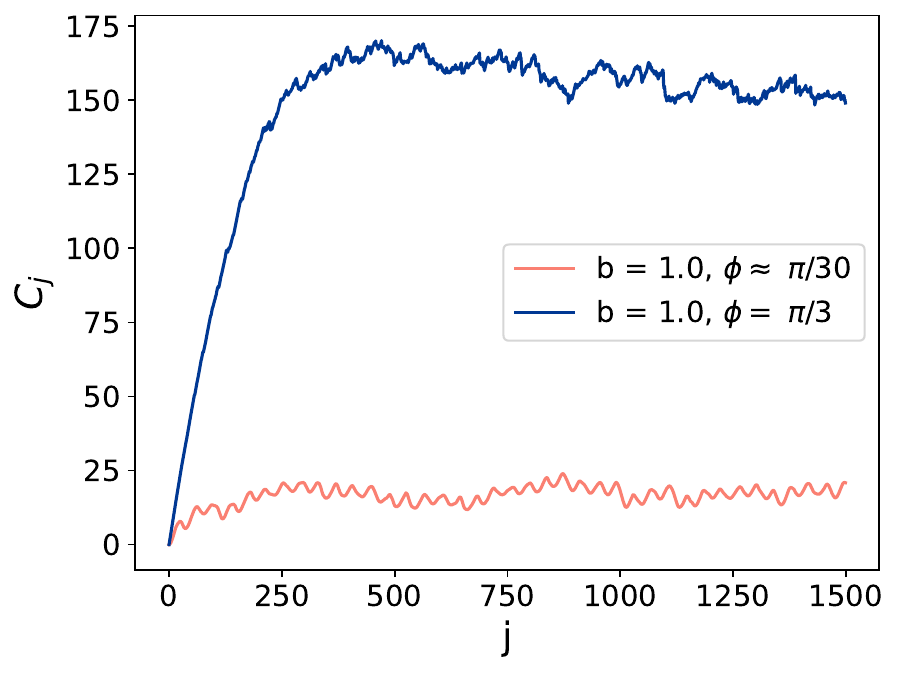}
    \includegraphics[height = 5.8 cm, width = 8.0 cm]{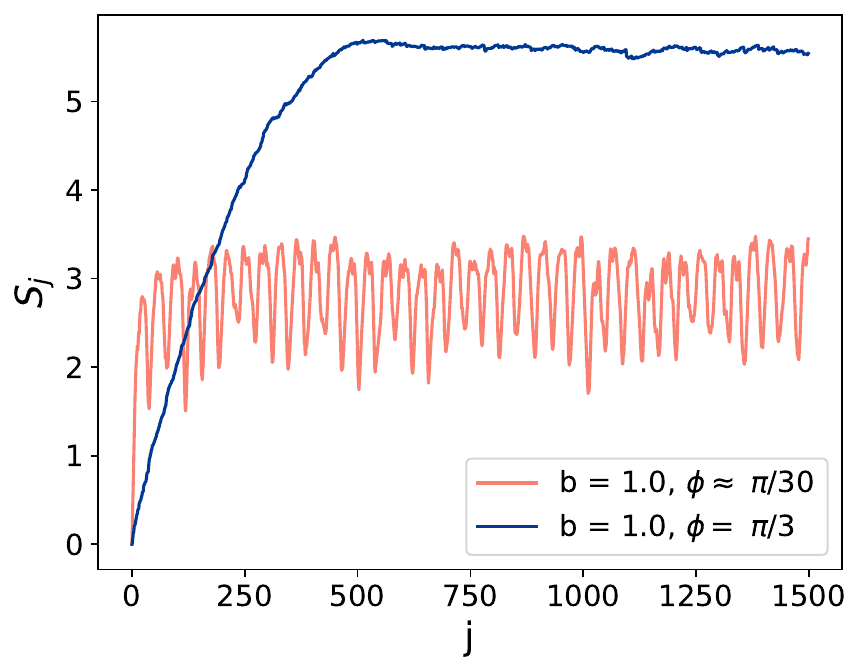}
    \caption{\small Spread Complexity (left) and Entropy (right) for different values of the coupling paramterised by $\phi$.}
    \label{fig:SC}
\end{figure}

Fig. \ref{fig:State} shows the behaviour of the Arnoldi coefficients $h_{n,n-1}$ and the real part of $h_{n,n}$ (the imaginary part has a similar behaviour).  The fluctuations in the Arnoldi coefficients are noticeably larger in the near-integrable case.
 The larger fluctuations signal the presence of localisation on the Krylov chain, a phenomenon that results in a suppressed value of K-complexity \cite{Rabinovici_2022, Dymarsky:2019elm}. This difference in the spread complexity saturation value in the two limiting cases is also clearly seen in figure \ref{fig:SC}. In the integrable limit we also observe that the spread entropy shows larger quasi-periodic fluctuations, and also its initial growth rate is faster although the growth is sustained for a longer time in the chaotic case.  It would be interesting to discern the physical mechanism underlying this behaviour
\footnote{Note that although the presented results are for a fixed value of the coupling constants and a particular initial state, we have checked that the above observations are generic and continue to hold for other choices.}. 

\subsection*{Dispersion of Arnoldi Coefficients}

We will now investigate the relation between chaotic dynamics and the dispersion of Arnoldi coefficients. The localisation length ($L_{loc}$) of the wavefunction is related to the dispersion or disorder: $L_{loc} \sim \sigma_h^{-1}$ \cite{fleishman1977fluctuations}. For this, we plot the variation in the standard deviation of Arnoldi coefficients as a function of the coupling parameter $\phi$.   We have scaled all the parameters such that the maximum value is set to one and the minimum value is set to zero. This means that a re-scaled variable is defined as
\begin{align}
X_s=\frac{X-X_{\text{min}}}{X_{\text{max}}-X_{\text{min}}}
\end{align}
analogous to the definition of the spectral parameter in eq. \eqref{etadefn}. Here $X$ can be the delocalisation parameter, or as discussed below, the time-averaged magnetisation or spread complexity saturation value. This facilitates comparison between all these quantities as the coupling is varied.

\begin{figure}[H]
    \centering
    \includegraphics[height = 5.7 cm, width = 8.0 cm]{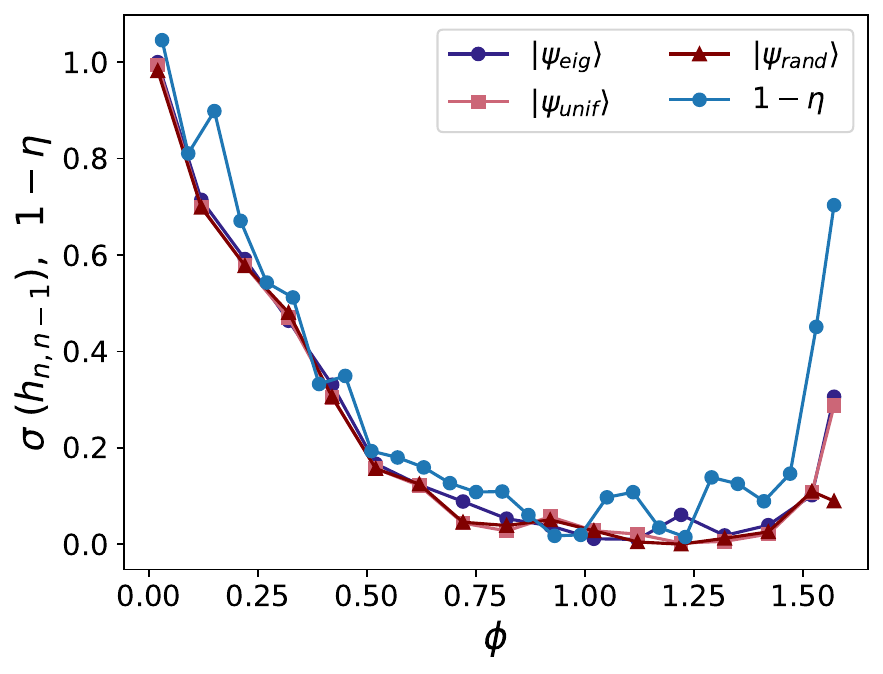}
    \includegraphics[height = 5.7 cm, width = 8.0 cm]{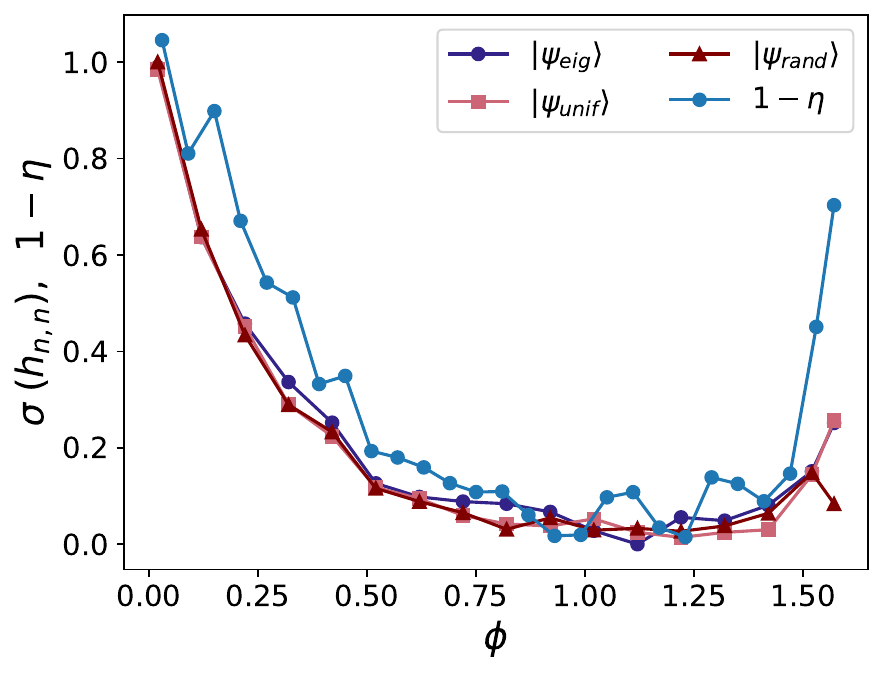}
    \caption{\small (rescaled) standard deviation of the Arnoldi coefficients $h_{n,n-1}$ and $h_n,n$ for different values of $\phi$. The variation in the spectral parameter is also shown for comparison.}

    \label{fig:dispersion}
\end{figure}

The different initial states chosen below include $\ket{\psi_{eig}}$ which is an $H_0$ eigenstate (all spins up). $\ket{\psi_{unif}}$ is the normalised state with the same constant coefficient as each of the component. In the third case of random initial states, the calculations were done for five different random initial states and averaged over for each value of $\phi$. It is noteworthy that different initial states have essentially overlapping dispersion plots \footnote{Besides the initial states and Arnoldi coefficients in fig.\ref{fig:dispersion}, we have checked this universality in the behaviour for several other cases as well. For autonomous spin chains, a similar observation for Lanczos coefficients was made in \cite{Scialchi:2023bmw}. In this reference some variants of the delocalisation length involving dispersion about the moving average and with the log ratio of Lanczos coefficients were also considered. However, the simple standard deviation suffices for our purposes to delineate the main features of the fluctuations in the Arnoldi sequences.}. We can infer that this dispersion or delocalisation length is a robust state-independent measure of chaotic dynamics.

\subsection*{Saturation value of spread complexity}

Another quantity of interest is the saturation value of spread complexity at late times \footnote{``Late" in this context means beyond the Heisenberg time scale.}. The definition we will use for the late-time spread complexity saturation value will be the discrete analogue of the one used in \cite{integrability_to_chaos}:
\begin{align}
    \overline{C_j} = \sum_{n=0}^{D_K-1}n \overline{|\psi_n|^2} \;\;\;\;\;\;\;\;\text{where, } \overline{|\psi_n|^2} = \lim_{m \to \infty} \frac{1}{m}\sum_{j=0}^{m} |\psi_n^{(j)}|^2 \label{saturation_value}
\end{align}

This saturation value is comparatively higher for chaotic system and this is related to delocalization on the Krylov chain. It has been considered as an indicator of chaos in recent literature \cite{integrability_to_chaos, Craps:2023rur, Nizami:2023dkf}. However, there are caveats related to the choice of initial states. Recent work \cite{Bernardo_2022, Scialchi:2023bmw} argued that the late-time saturation values for the choice of certain initial states does not always agree with the other chaos measures such as r-statistics. 

\begin{figure}[H]
    \centering
    \includegraphics[height = 5.8 cm, width = 8.0 cm]{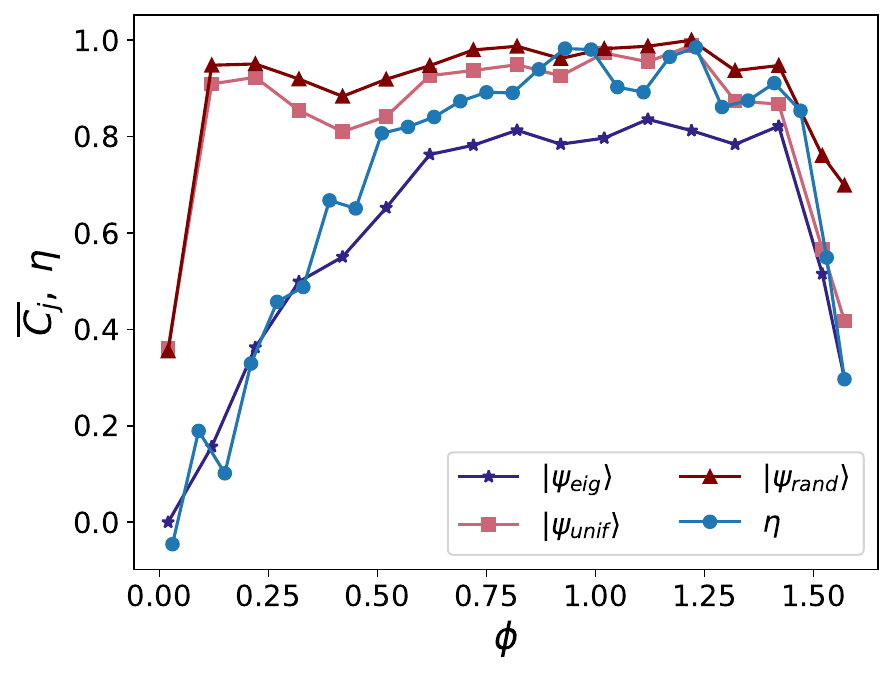}
    \caption{\small Rescaled saturation value of the spread complexity as a function of the coupling parameter $\phi$ for different initial states.}
    \label{fig:saturation_value}
\end{figure}

 The saturation value, for the same set of initial states as before, is plotted below as a function of the coupling $\phi$.  We notice that when the initial state is an eigenstate of $H_0$, the results match with the integrability to chaos transition as shown by the $\eta$ parameter in fig. \ref{fig:saturation_value}. However the same is not always true for other initial states. 

\subsection{Magnetisation}
The mean magnetisation is a classic order parameter quantifying the disorder in the many-body state for a system.  We will compute the time-dependent expectation value of the operator  $J_z=\sum_i \sigma^z_i$  with the initial state being all spins up, and also the variation of the time-averaged value $\overline{J_z}$ as the parameter $\phi$ is changed. The long-time average $\overline{J_z}$ serves as an order parameter for the sytem and is defined as:
\begin{align}
    \overline{J_z} = \lim_{T \to \infty} \frac{1}{T} \sum_{j=0}^T \bra{\psi_j} J_z \ket{\psi_j}
\end{align}
where the state $\ket{\psi_j}$ is defined as $\ket{\psi_j}=U_F^j \ket{\psi_0}$.

\begin{figure}[H]
    \centering
    \includegraphics[height = 5.8cm, width = 8.0 cm]{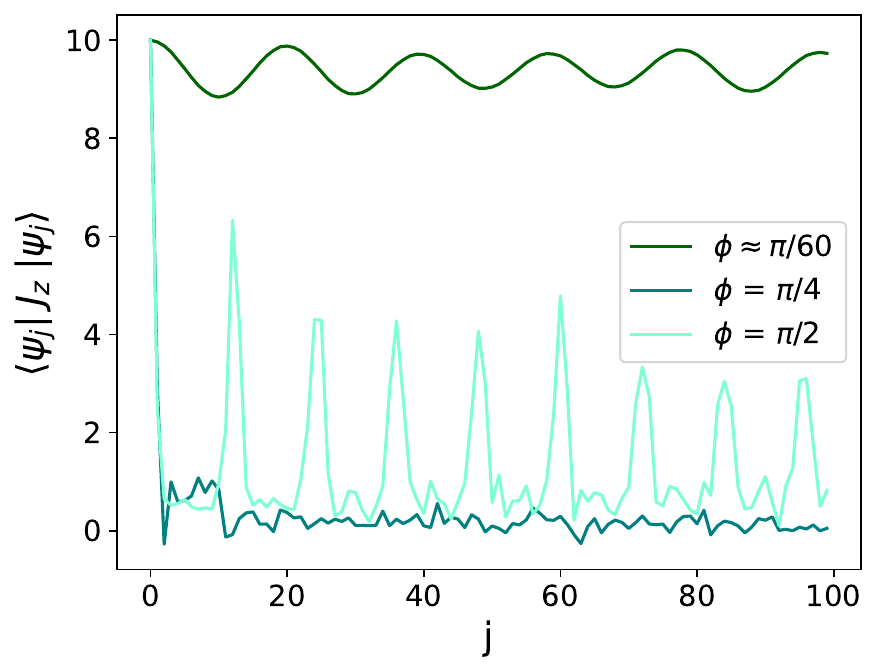}
    \includegraphics[height = 5.8cm, width = 8.0 cm]{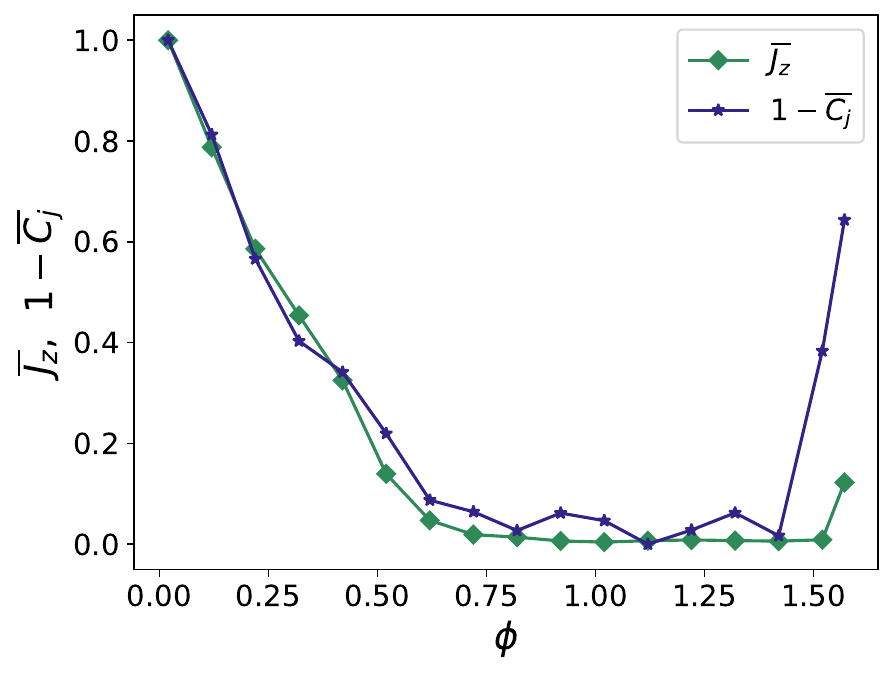}
    \caption{\small Evolution of expectation value of the total spin operator $J_z$ (left) and its time average (rescaled) for different values of parameter $\phi$ (right), with the averaged spread complexity also shown for comparison.}
    \label{fig: Magnetization}
\end{figure}

As shown in the fig. \ref{fig: Magnetization} (left) the mean magnetisation shows a periodic variation with time and a large average value in the weakly coupled case, as expected in this ferromagnetic case with a small amount of disorder. However for the chaotic case, the magnetisation drops to near-zero quickly. For the integrable but strongly coupled case, there are quasi-periodic fluctuations. The figure on the right
shows the behaviour of the (rescaled) order parameter $\overline{J_z}$ and the time averaged spread complexity $\overline{C_j}$ discussed in the previous section. For a (time-independent) spin-chain described by the LMG model, results for the magnetisation were obtained recently in
\cite{Bento:2023bjn}. There it was shown analytically that when the initial state is an eigenstate of the pre-quench hamiltonian ($H_0$), then $\overline{J_z}$ and $\overline{C(t)}$ are the same upto an additive constant. This result essentially followed from the observation that the Lanczos basis is the same as the pre-quench energy eigenbasis. As shown in Fig. \ref{fig: Magnetization} (right), where the initial state is an $H_0$ eigenstate, $\overline{J_z}$ and $\overline{C_j}$ have similar behaviour although in this case the Krylov-Arnoldi basis doesn't match with the $H_0$ eigenbasis.

\subsection{Non-local Kicked Ising Chain}
We will now turn to a variant of the above kicked Ising spin chain model where the interactions are maximally non-local. In particular, the Hamiltonian contains all-to-all spin interactions with uniform coupling.  

\begin{align}
    H = H_{\text{local}} + \gamma \sum_{i<j} \sigma_i^z \sigma_{j}^z
\end{align}
where, $H_{local}$ is the Hamiltonian given in Eq. \eqref{LSC}. The Floquet operator for non-local interactions is given by:
\begin{align}
    U_F = \exp{ -\text{i}\, \gamma \sum_{i<j}\sigma_i^z \sigma_{j}^z} U_{F\,\text{(local)}}
\end{align}
where, $U_{F\,\text{(local)}}$ is the Floquet operator given in Eq. \eqref{FSC}. For time-independent spin-chains, the effects of non-locality on (operator) complexity have been recently studied in
\cite{bhattacharya2023krylov}.

Even without the periodic drive, such systems with all-to-all two-body interactions are known to be {\it fast scramblers} \cite{Sekino:2008he, Lashkari:2011yi, Belyansky:2020bia}. This means that the exponentially fast spreading of quantum information in such systems involves a time scale, the scrambling time $t_s$, that is logarithmic in the system size $N$. Other examples of fast scramblers are brownian quantum circuits, SYK model and black holes.

We will study the effect of non-locality on spread complexity in the non-trivially integrable regime at $\phi = \pi/2$. This system can be solved via Jordan-Wigner transformation \cite{Lieb:1961fr}. At the value of $\phi=\pi/2$, the system has a parity symmetry as well as Z-reflection symmetry. 


\begin{figure}[H]
    \centering
    \includegraphics[height = 5.8 cm, width = 8.0 cm]{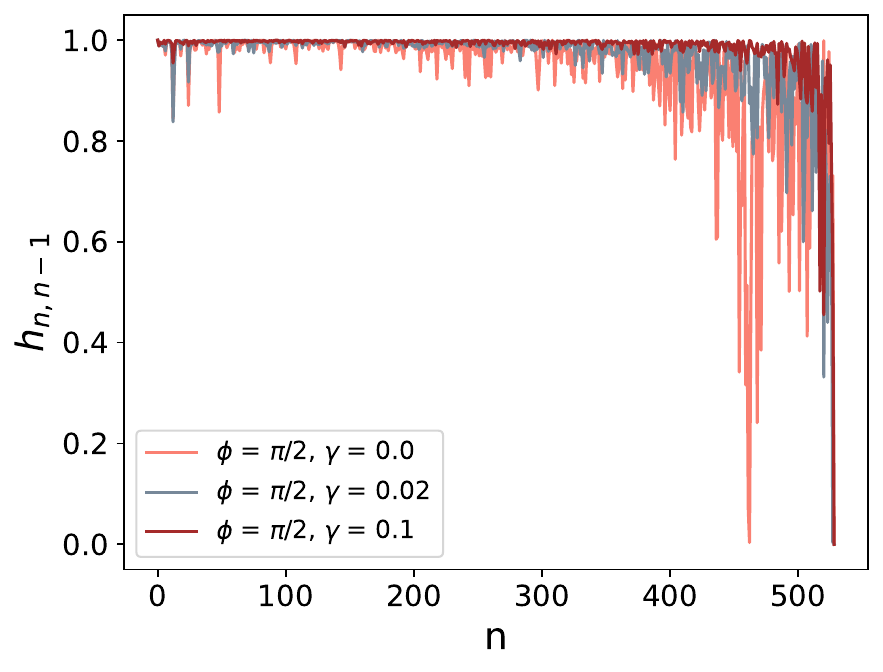}
    \includegraphics[height = 5.8 cm, width = 8.0 cm]{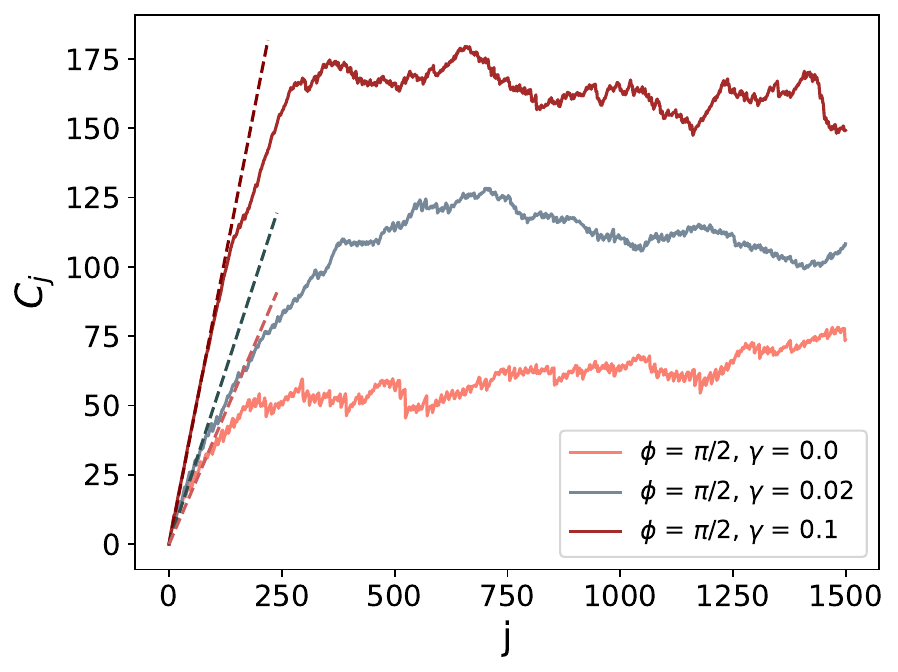}
    \includegraphics[height = 5.8 cm, width = 8.0 cm]{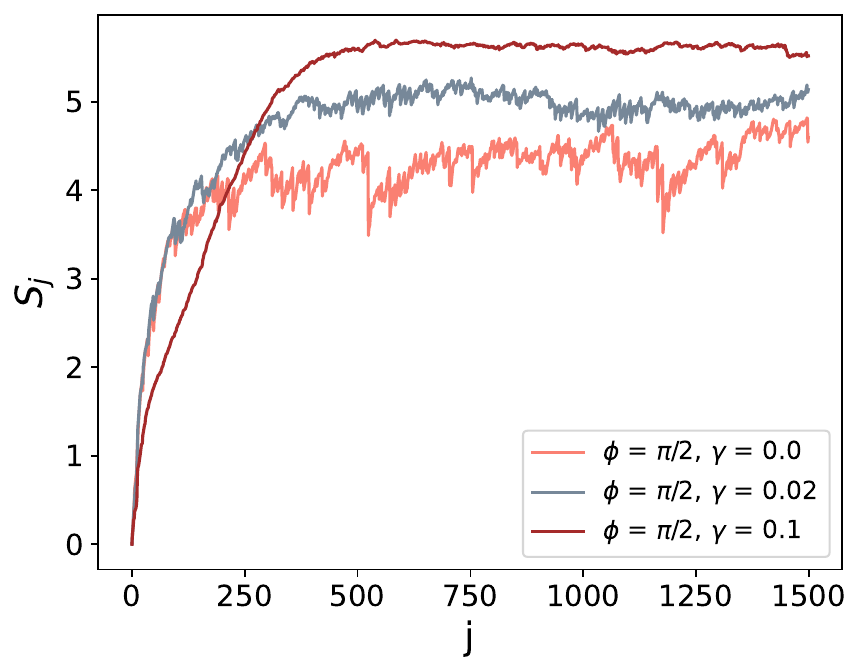}
    \includegraphics[height = 5.8 cm, width = 8.0 cm]{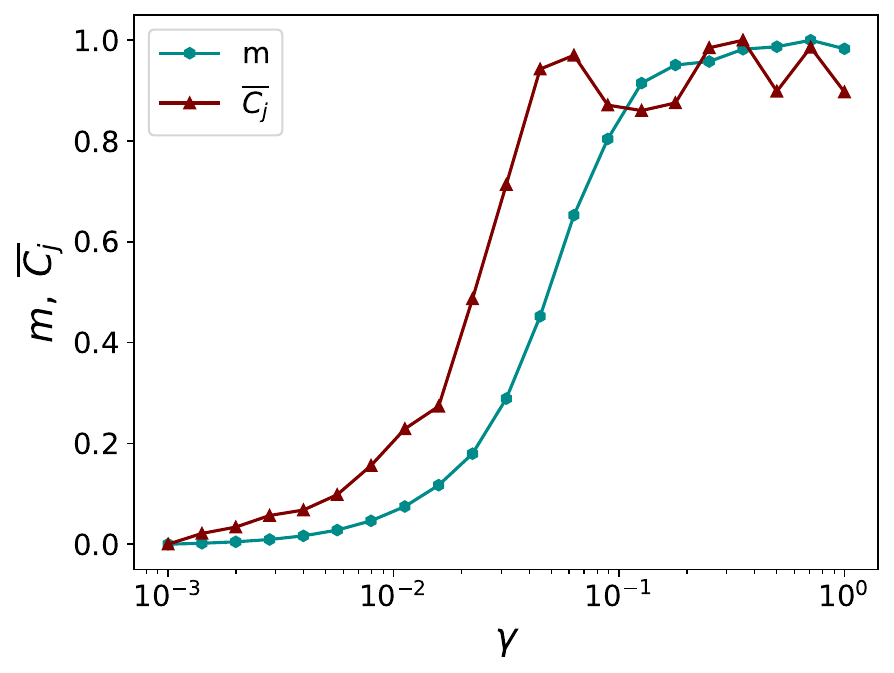}
    \caption{\small Arnoldi coefficients (top left), state complexity (top right), entropy (bottom left) for the Ising chain with local and non-local interactions in the integrable and chaotic regime. The bottom right figure shows how the (normalised) averaged spread complexity and initial slope grows as the non-local coupling is increased.}
    \label{fig:Arnoldi_non-local}
\end{figure}

The effect of turning on the non-local coupling ($\gamma$) is to decrease the disorder in the Arnoldi coefficients, as made clear by Fig. \ref{fig:Arnoldi_non-local}. This figure also clearly shows how the saturation value of the spread complexity  and its initial growth rate increases with $\gamma$. This makes manifest that non-locality of the interaction induces chaos even at relatively small coupling constants. Given the close relationship between locality and integrability \cite{Wanisch:2022gyr}, this is not unexpected. However, it is interesting that even perturbatively small {\it non-local} couplings can open the way for maximal chaos. Note that (fig.\ref{fig:Arnoldi_non-local}, top right) turning on $\gamma$ increases the saturation value of the spread complexity (due to Krylov chain delocalisation, a signature of chaos), and also increases the initial growth rate of spread complexity (signaling a faster scrambling in the system due to non-locality). Fig.\ref{fig:Arnoldi_non-local} (bottom right)  also shows the variation of the slope ($m$) for the initial growth rate of spread complexity as a function of the non-local coupling $\gamma$. We observe that the bound ($m<1$) discussed in section \ref{KCP} is obeyed. We also find that $\gamma$ has no substantive effect when it is turned on with the system already in the chaotic regime. 
  
\subsection{Tuning the driving frequency}
\begin{figure}[H]
    \centering
    \includegraphics[height = 5.8 cm, width = 8.0 cm]{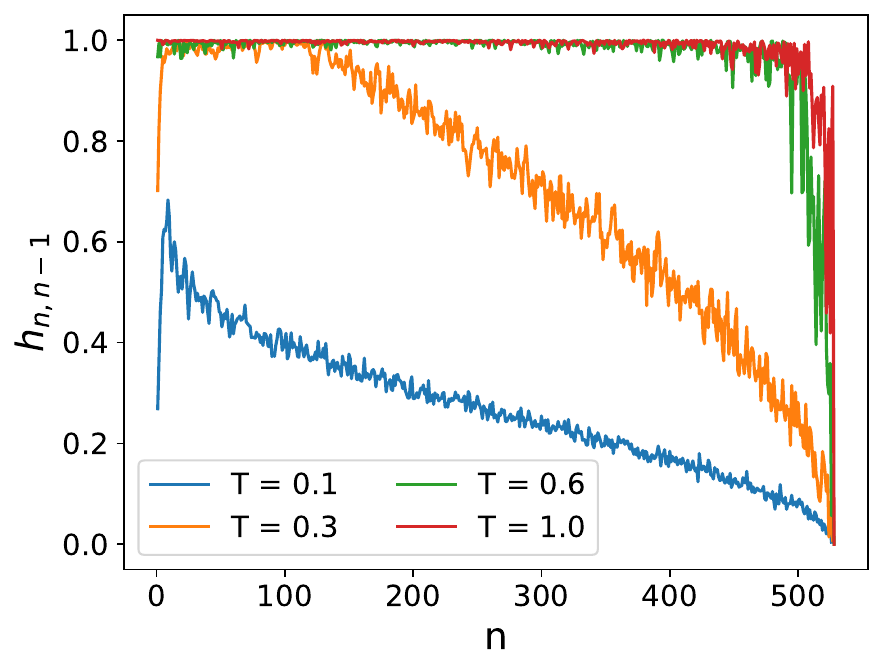}
    \includegraphics[height = 5.8 cm, width = 8.0 cm]{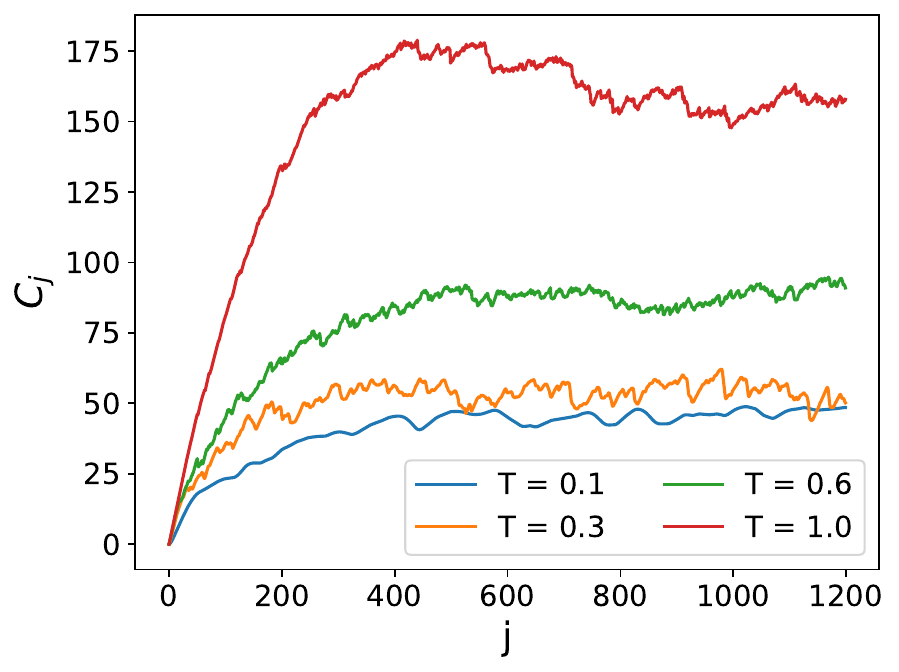}
    \caption{\small Arnoldi coefficients (left) and state complexity (right) for the kicked Ising spin-chain for different values of kicking period $T$. In the high frequency limit, note the Lanczos-like behaviour of the Arnoldi coefficients and the suppression in the spread complexity saturation.}
    \label{fig: t_dependence}
\end{figure}

In our analysis so far, the time period of driving was set to unity. We will now investigate the effects of varying the driving frequency on the Arnoldi coefficients and spread complexity. The graphs below shows the Arnoldi coefficients and spread complexity for different values of time period $T$ in the chaotic kicked Ising spin-chain. We can see that in the high frequency (small $T$) limit, the Arnoldi coefficients show the signature behavior of Lanczos coefficients -- initial linear growth followed by a descent \cite{Rabinovici:2019wsy}. In appendix B we show how the Arnoldi Krylov construction goes over to the Lanczos one, in the limit of a high frequency drive. The saturation value of the spread complexity also decreases as the time-period of kicking decreases. This result may be explained as follows. For a fixed drive frequency and varying coupling constant, the integrable case has extra symmetries characterised by the presence of conserved quantities. Thus in the integrable limit, level crossings and clustering leads to degeneracies and quasi-degeneracies in the spectrum. This results in enhanced disorder in the Lanczos sequence resulting in a suppressed complexity saturation value (the mechanism for this Krylov localisation is explained in \cite{Rabinovici_2022}) . Such conserved quantities can also arise in the high frequency limit of the periodic drive. In this limiting case,
\begin{align}
U_F&=\exp (-i T H_0) \exp (-iTV) \nonumber \rightarrow \exp \big(-i T (H_0+V)\big)=\exp(-iT H_F)
\end{align}
and thus operators commuting with the effective Floquet hamiltonian $H_F$ are conserved in the high frequency limit. As seen in fig. \ref{fig: t_dependence}, in this case we do not have enhanced disorder in the Arnoldi coefficients in the high frequency limit. Nevertheless we do have a suppressed complexity saturation value which is thus not due to the usual Krylov localisation, as in the in the near integrable case when the couplings are tuned to smaller values. 

\begin{figure}[H]
    \centering
    \includegraphics[height = 5.8 cm, width = 8.0 cm]{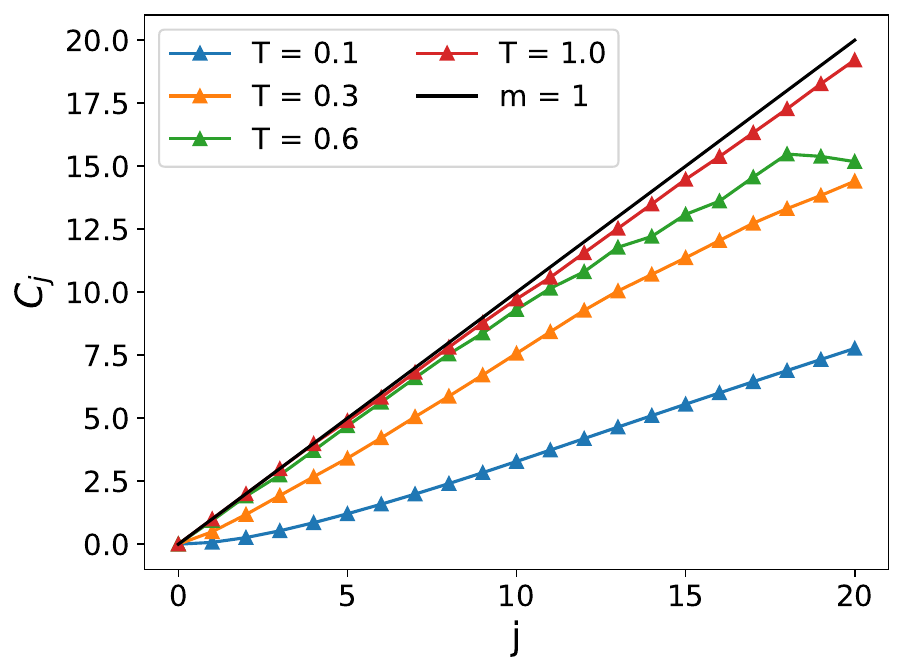}
    \includegraphics[height = 5.8 cm, width = 8.0 cm]{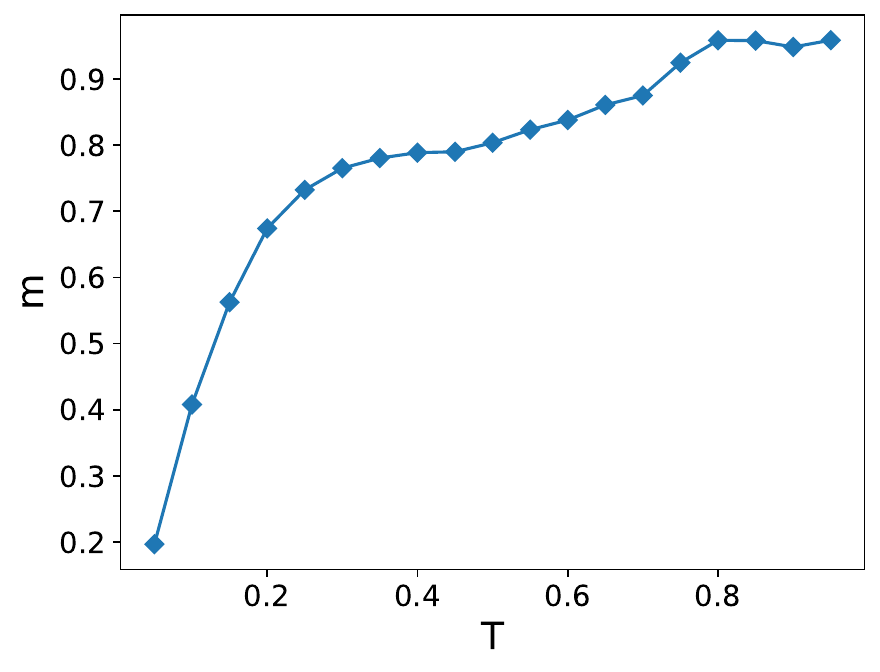}
    \caption{\small Initial growth of Arnoldi coefficients (left) and slope of the best fit line for different values of kicking period $T$ (right).}
    \label{fig: t_dependence slope }
\end{figure}

We also zoom into the region of initial linear growth and compute the slope as a function of the time-period of the drive. In fig. \ref{fig: t_dependence slope } the bound on the slope is represented by the black line. Increasing the time period, the initial growth rate increases and tends towards the maximal possible as shown in fig. \ref{fig: t_dependence slope } (right). These observations are in accord with the result of section \ref{rev}, where we showed that Krylov chain dynamics implies that the initial growth rate of the spread complexity is at most linear (with slope 1).

\section{Discussion} \label{dis}

We have studied the dynamics of a kicked Ising spin-chain (local as well as non-local) through a number of chaos quantifiers. The utility of the various Krylov measures is highlighted by the accord with spectral quantifiers of non-integrable dynamics. In section \ref{ising} we studied the regular to chaotic transition in the kicked Ising spin-chain resulting from tuning its coupling strength. A noteworthy result here was the concurrence between several different diagnostics of non-integrable dynamics namely the spectral parameter $\eta$, the delocalization parameter $\sigma_h$, the saturation value of spread complexity $\overline{C_j}$ and the average magnetization $\Bar{J_z}$. As noted in the literature on autonomous systems \cite{Scialchi:2023bmw, Bernardo_2022}, we also find that the saturation value of complexity has dependence on the initial state, although the match with the spectral parameter is good for an initial state that is localised in the $H_0$ eigenbasis. 

Our results in this section also show that maximal chaos with fast scrambling can arise in kicked spin-chains due to the presence of non-local interactions, even with weak coupling strengths (besides the usual route to chaos due to strong local couplings).  We also discussed the effect of changing the frequency of the drive and saw that, in the high frequency limit, we obtain a suppressed saturation value of the spread complexity and the dynamics of Arnoldi coefficients reduces to the Lanczos case. We reiterate that this suppressed saturation value is not due to the usual Krylov localisation. Even though the Arnoldi sequences do not show larger disorder, nevertheless the saturation value is suppressed.

One can further study the transition from integrability to chaos in harmonically driven or periodically quenched spin-chains utilising the methods used in this paper. Also, in all these models it would be interesting to investigate the existence of dynamical quantum phase transitions in the thermodynamic limit via non-analyticities in the Loschmidt amplitude and K-complexity saturation \cite{Heyl:2013ywe,Heyl:2017blm, Bento:2023bjn}.

\section*{Acknowledgements}
We thank Arul Lakshminarayan, Adolfo del Campo, Irfan N. Mir and Philip Cherian for useful discussions. AAN thanks Anatoly Dymarsky for useful comments on the manuscript. AWS would like to thank the Research and Development Office, Ashoka University and Axis Bank for financial support. AWS  would also like to thank the participants of the workshop ``Quantum Dynamics and Chaos: Modern Perspectives" (Ashoka University, March 9-11, 2024) where some of the results of this paper were presented. We would like to acknowledge use of the physics department server and the QuSpin package \cite{Weinberg:2017igw} for the numerical calculations. We are also thankful to the referees for their questions and suggestions.

\appendix

\section*{Appendix A: Special case}
\begin{figure}[H]
    \centering
    \includegraphics[height = 5.8 cm, width = 8.0 cm]{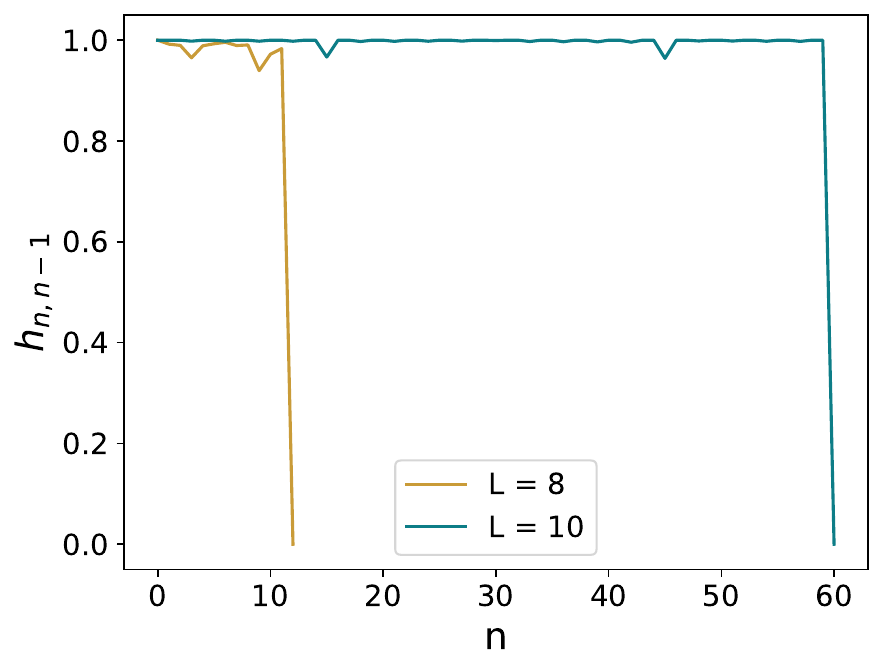}
    \includegraphics[height = 5.8 cm, width = 8.0 cm]{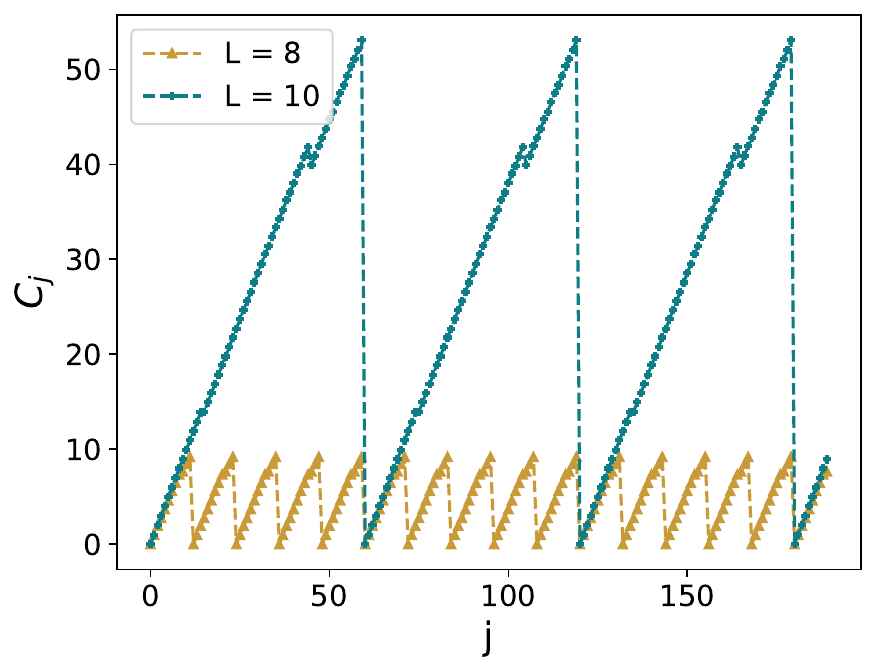}
    \caption{\small Arnoldi Coefficients and Spread complexity in the case of resonance in KIC.}
    \label{fig: resonance}
\end{figure}

We will look at an interesting special case \footnote{We thank Arul Lakshminarayan for bringing this case to our attention.} when the coupling constants take on particular values in the kicked Ising chain. We start with a modified form of the Floquet operator
\begin{align}
    U_F = \exp{ -i h_x \sum_{i=1}^N \sigma_i^x } \exp{ -i J \sum_{i=1}^{N-1} \sigma_i^z \sigma_{i+1}^z  -i h_z\sum_{i=1}^N  \sigma_i^z}
\end{align}
and work with the values of parameters: $h_x = h_z = J=\pi/4$. This is the particular case of the self-dual kicked Ising spin chain also known as a dual unitary model 
\cite{Akila:2016ltx, Bertini:2018wlu}. In such cases, $U_F^p$ becomes the identity operator for some integer $p$ which depends on the number of spins. A similar observation was made in the case of Clifford circuits in \cite{suchsland_krylov_2023}. Note in fig.\ref{fig: resonance} the linear growth (with slope 1) and recurrence in the spread complexity. As noted in section \ref{KCP} the $h_{n,n-1}$ coefficients being unity leads to a maximal (linear growth) of spread complexity, which is what this example illustrates. The recurrence also has a period that scales with the system size. This special case was also discussed in \cite{Pal:2018nvj} where the operator entanglement was studied.


\section*{Appendix B: High frequency limit for the Floquet operator}\label{HFL}
In most of the calculations in this paper, we have set the time-period of delta-function kicks to unity. However, we can study the behavior of Arnoldi coefficients and spread complexity by changing the frequency of the kick. The general Floquet operator is given as:
\begin{align}
    U_F = e^{-iH_0 T} e^{-iVT} = e^{-i(H_0 + V)T - \frac{1}{2}[H_0,V]T^2 + \dots}
\end{align}
where the second equality comes from BCH relation. In the limit that $T \to 0$, we can approximate the Floquet operator upto the first order as: 
\begin{align}
    U_F \approx e^{-i(H_0 + V)T} = e^{-iH'T}
\end{align}

\noindent If we plug this expression in the Arnoldi iteration:
\begin{align}
   \ket{A_n}  &= e^{-iH'T}\ket{K_{n-1}} - \sum_{j=0}^{n-1} \bra{K_j}e^{-iH'T} \ket{K_{n-1}} \ket{K_j} 
\end{align}
and take the terms upto the first order in Taylor expansion  of the exponential:
\begin{align}
    \ket{A_n}  = \ket{K_{n-1}} - iTH'&\ket{K_{n-1}} - \sum_{j=0}^{n-1} \biggl[\braket{K_j|K_{n-1}} \nonumber \\ &- iT \bra{K_j}H'\ket{K_{n-1}} \biggr]\ket{K_j} \nonumber \\
    = - iTH'\ket{K_{n-1}}  + &\sum_{j=0}^{n-1}  iT \bra{K_j}H'\ket{K_{n-1}} \ket{K_j} \nonumber 
\end{align}
The action of the Hamiltonian in the Krylov Basis is:
\begin{align}
    H'\ket{K_{n-1}} = a_{n-1}\ket{K_{n-1}} + b_n\ket{K_n} + b_{n-1} \ket{K_{n-2}}
\end{align}
Plugging this in the above expression and using the orthogonality relation for Krylov Basis vectors:
\setlength{\jot}{10 pt}
\begin{align*}
    \ket{A_n} = - iTH'\ket{K_{n-1}}  + \sum_{j=0}^{n-1}  iT \biggl[a_{n-1}\braket{K_j|K_{n-1}}  \nonumber \\ + b_n\braket{K_j|K_n} + b_{n-1} \braket{K_j|K_{n-2}}\biggr] \ket{K_j} \nonumber \\
    = - iT [H'\ket{K_{n-1}} - a_{n-1}\ket{K_{n-1}} - b_{n-1} \ket{K_{n-2}}] 
\end{align*}
which, upto a constant factor, is the exact expression for Lanczos algorithm. Thus, we expect the behavior of the Arnoldi coefficients to show features of Lanczos coefficients in the high frequency limit $T \to 0$. 

\section*{Appendix C: Integrable to chaos transition in kicked Bose-Hubbard dimer}\label{BH}

Another interesting many-body system that shows an integrable to chaos transition is the kicked Bose-Hubbard dimer model. Recently, \cite{Liang:2023xaj} explored the transition in this model using spectral statistics, Husimi distribution, OTOC and entanglement entropy. This appendix will use spread complexity as another dynamical measure to understand the integrable or chaotic nature of this bosonic system.
\par \medskip
The two-site model is described by the Hamiltonian \cite{Liang:2023xaj}:
\begin{align}
    H = \nu(b_0^\dagger b_1 + b_1^\dagger b_0) + \frac{U}{N}\sum_{i=0}^1 n_i(n_i-1) + \frac{\mu}{2} (n_0 - n_1) \delta_T(t)
\end{align}
where, $\nu$ is the hopping amplitude between the sites, $U$ sets the on-site interaction strength, and $N$ is the total number of particles. Similarly, $b_i$ and $b_i^\dagger$ are bosonic annihilation and creation operators and $n_i = b_i^\dagger b_i$ is the number operator at site $i$. They satisfy the usual commutation relations:
\begin{align}
    [b_i, b_j] = \delta_{ij}, \;\;\;\; [b_i^\dagger, b_j^\dagger] = [b_i,b_j] = 0
\end{align}

The parameter $\mu$ in the Hamiltonian controls the periodically applied potential difference between the sites with time-period $T$. The model can be represented in terms of the angular momentum operators:
\begin{align}
    H = 2 J_x + \frac{k}{N} J_z^2 + \mu J_z \,\delta_T(t)
\end{align}
where the $J$ operators are mapped to bosonic creation and annihilation operators as:
\begin{align}
    J_x = \frac{b_0^\dagger b_1 + b_1^\dagger b_0}{2},\;\;
    J_y = \frac{i (b_0^\dagger b_1 - b_1^\dagger b_0)}{2},\;\;
    J_z = \frac{b_0^\dagger b_1 - b_1^\dagger b_0}{2}
\end{align}
These operators satisfy the commutation relation $[J_i, J_j] = i\epsilon_{ijk}J_k$ and $J^2 = \frac{N}{2}(\frac{N}{2}+1)$ is a conserved quantity. Here, $\nu$ is set to one, and $k = 2U$ \cite{Liang:2023xaj}. The corresponding Floquet operator is:
\begin{align}
    U_F = \exp{-i\mu J_z } \exp{-i \left(2J_x + \frac{k}{N}J_z^2 \right)}
\end{align}

\par \medskip
As before, we will study the late-time saturation value of spread complexity and size of fluctuations in Arnoldi coefficients, and compare it with the spectral statistics parameter $\eta$. We will take the eigenstate of the $J_x$ operator as the initial state and a system size $J = 250$ for the calculations.

It was shown in \cite{Liang:2023xaj} that the system shows integrable to chaos transition for $\mu = 3.0$ while remains integrable for $\mu = 6.0$. We can see that the measures from spread complexity follows a similar trend -- standard deviation of Arnoldi coefficients gets suppressed and late-time saturation value of spread complexity increases during the transition to chaos for $\mu =3.0$. The values fluctuate around a mean value for $\mu = 6.0$ as expected for a system that remains integrable. 

\begin{figure}[H]
    \centering
    \includegraphics[height = 5.8 cm, width = 8.0 cm]{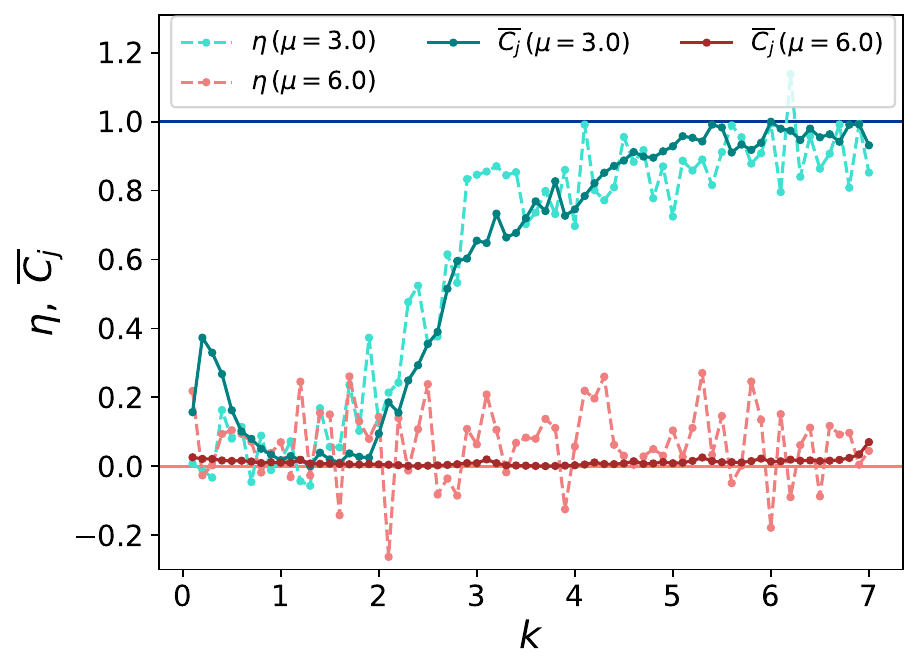}
    \includegraphics[height = 6.1 cm, width = 8.0 cm]{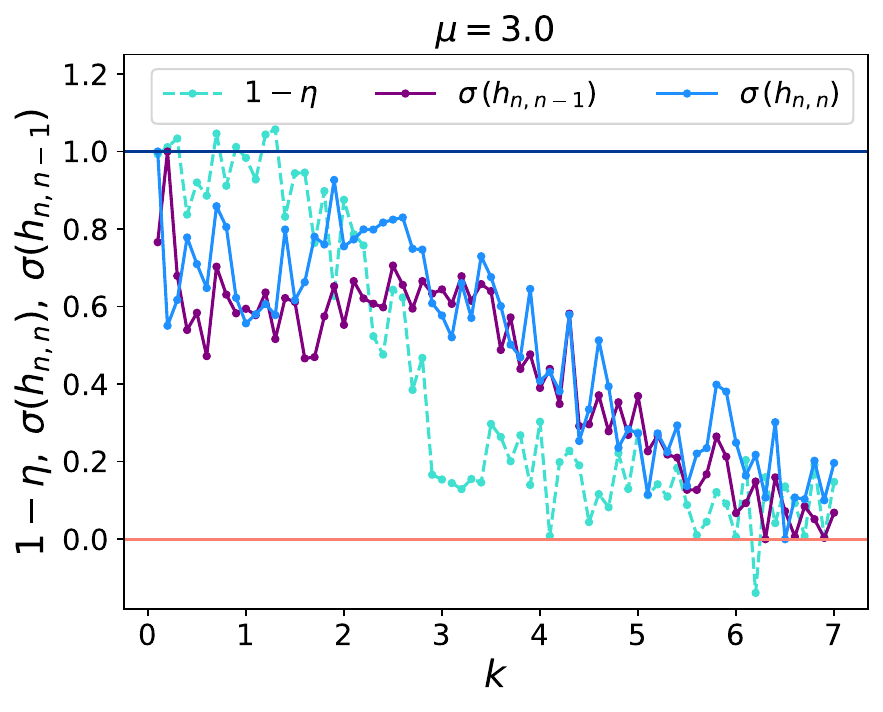}
    \includegraphics[height = 6.1 cm, width = 8.0 cm]{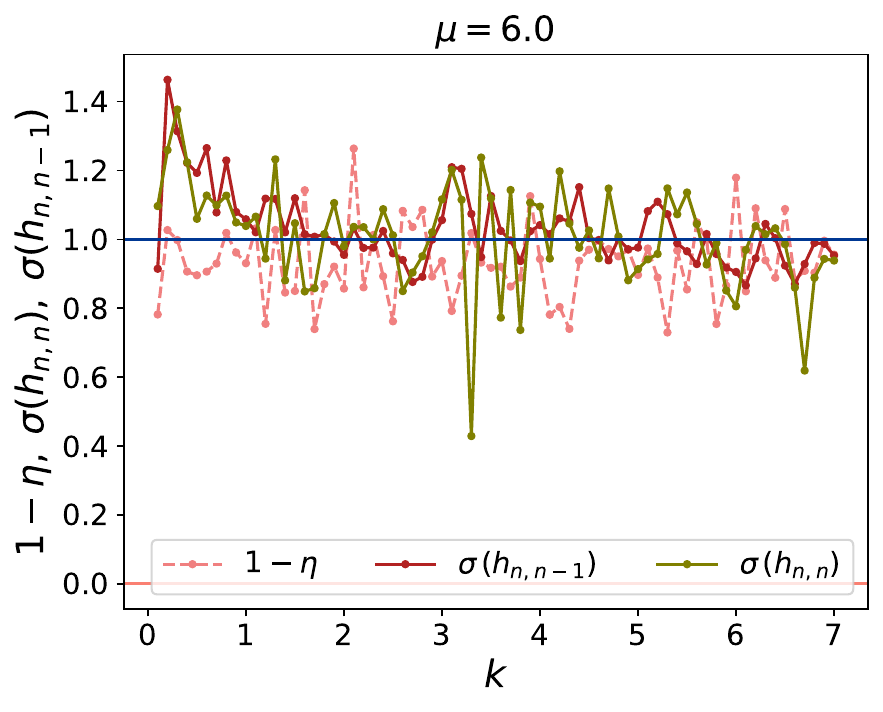}
    \label{fig:Arnoldi_non-local }
    \caption{\small Late-time saturation value of the spread complexity, standard deviation of the Arnoldi coefficients, and parameter $\eta$ for different values of coupling parameters $\mu$ and $k$.}
\end{figure}

\printbibliography
\end{document}